\documentclass[a4paper]{article}
\usepackage[dvips]{graphicx}

\begin{document}
\title{\bf CHANGE OF PRIMARY COSMIC RADIATION NUCLEAR COMPOSITION
IN THE ENERGY RANGE  $10^{15}  -  10^{17}$  eV }

\author{T.T.~Barnaveli\thanks{e-mail: mantex@caucasus.net,~~~~
barnaveli@hotmail.com}\hspace*{2mm}(1),
~T.T.~Barnaveli (jr) (1),~ A.P.~Chubenko (2), \\
N.A.~Eristavi (1),~I.V.~Khaldeeva (1),~N.M.~Nesterova (2)~ \\
  and Yu.G.~Verbetsky (1) \\   \\
(1) {\small\it Tbilisi Institute of Physics,}\\
{\small\it Georgian Acad.Sci., Tamarashvili 6, Tbilisi 380077, Georgia} \\
(2) {\small\it Lebedev Physical Institute,} \\
{\small\it Russian Acad.Sci., Leninsky prosp.53,Moscow 101000, Russia}}
\date{}

\maketitle

\begin{abstract}
The dependence $E_h (N_e)$ of Extensive Air Shower (EAS) hadronic
component energy flux on the number $N_e$ of particles in EAS is
investigated in the primary energy  range of the order of
$10^{15} -  10^{17}$  eV. The work was aimed at checking the
existence of irregularities of  ~$E_h (N_e)/N_e$   behavior at
these energies in several independent experiments.

The investigation is carried out using large statistical material
obtained at different configurations of experimental apparatus
and under different triggering conditions.

The existence of irregularities of  $E_h (N_e)/N_e$ behavior in
the region $N_e  > 2*10^6$  is confirmed. These irregularities
have the character of sharp deeps and are located near the same
values of $N_e$  regardless of the experimental material and
selection conditions used. So, at recent stage of research the
existence of these irregularities of  $E_h (N_e)/N_e$ behavior in
the range  of $N_e    >2*10^6$   may be regarded as reliably
established. This fact supports our earlier conclusion on the
existence of primary cosmic radiation (PCR) nuclei spectra cutoff
effect in the primary energy region  $10^{15}  -  10^{17}$  eV
\cite{1,4}.
\end{abstract}

\section{\bf Introduction}
\vspace{1pt}

~~~In  \cite{1} - \cite{5} we quoted some results which enable to
suppose the existence  of the PCR nuclei spectra cutoff in the
energy range $10^{15}  -  10^{17}$  eV.  In \cite{1} the analysis
was based on the investigation of high energy muon groups in EAS.
The sharp decrease of the energy and intensity of high energy muon
groups in EAS of primary energy higher than $2*10^{16}$  was
discovered. In \cite{2,3} and \cite{4} the fluxes of the EAS
hadronic component energy $E_h$ were analyzed . The essential
irregularities of the type of strict deeps in hadronic component
energy fluxes were discovered in the region of EAS particle
numbers $N_e > 2*10^6$, corresponding to the primary energies $E_0
> 5*10^{15}$ eV.

Analysis of this phenomenon  leads to the conclusion of the
existence  of PCR nuclei spectra cutoff in the primary energy
region  $10^{15}  -  10^{17}$  eV~  \cite{1} - \cite{4}.  In
\cite{1} and \cite{3} the interpretation of this phenomenon was
given - it can be explained by the destruction of PCR nuclei on
some monochromatic background of interstellar medium, consisting
of the light particles of mass $< 30$ eV.

The aim of this work is to carry out the investigation of EAS
hadronic component energy flux using the additional statistical
material, obtained at different configurations of the
experimental apparatus and at different triggering and selection
conditions. Thereby the aim was  to compare the results of
independent experiments.

Usage of essentially improved algorithms of data handling and
increased capacity of computers made it possible to establish the
EAS parameters with sufficiently high accuracy and to set much
more rigid conditions at selecting the events in each separate
case. This in turn enabled us to extract efficiently the
peculiarities in $E_h$($N_e$) behavior and increase the accuracy
of their localization along the $N_e$  axis. The above mentioned
earlier results and conclusions were confirmed with high
reliability.

\section{\bf Installation and experimental material }

~~~The archive material obtained by means of the hadron
calorimeter of Tian - Shan high mountain installation was used for
the analysis. The Tian - Shan EAS complex installation is located
at the height of 3335 $m$ above sea level.  The detailed
description of the installation can be found in \cite{6}. The data
bank was created in 1980 and its description is provided in [7].
The information concerning the EAS parameters was obtained by
means of the part of the installation consisting of a central
"carpet" of scintillators (64 x 0.25 $m^2$ ), of 8 groups of
scintillators with a  total area of 22 $m^2$ , situated
symmetrically at the distances of 15 and 20 $m$ from the center of
installation and of a group of scintillators with a  total area of
10 $m^2$  at the distance of 73 $m$ from the center of
installation. The accuracy of the EAS coordinate estimation rises
with the increase of EAS age parameter and of $N_e$ .  At $N_e =
10^5$ the error is approximately 0.6 - 0.7 $m$  \cite{8}. Taking
into account that in our consideration the most important region
of $N_e$ is higher than $4*10^5$ , the corresponding error will
not exceed the above mentioned value. The energy flux of EAS
hadronic component was measured by means of a multilayer
ionization calorimeter with the area of each layer 36~$m^2$. At
different stages of the experiment the calorimeter contained from
16 to 19 layers of ionization chambers with the total number of
chambers from 768 to 912 correspondingly. Between the layers of
chambers lead filters of the total thickness $1050~ g/cm^2$  were
situated.

 As is well known the data obtained with the ionization calorimeter
at the energies under consideration do not allow for
distinguishing the separate hadrons, especially in the central
parts of EAS.  One can judge about the number of particles and
about their individual energies only approximately, by indirect
methods. Therefore we analyze immediately the energy fluxes of
hadrons inside the fixed circle of a fixed radius around the
center of EAS as a function of the particle number $N_e$ in EAS or
of the primary energy $E_0$ .

 The importance of the precise measurement of the distance from
the hadron energy registration point to the EAS axis is obvious.
Consequently the analysis was focused on the investigation of the
energy flux through each separate chamber of the calorimeter in
each event. For each single chamber the distance from its
geometrical center to the shower axis was measured taking into
account the axis inclination angle. The whole interval of
investigated distances was divided into sections of the $R ~cm$
length. $R=20 ~cm$ when working inside the $14 ~m$ diapason of
core distances (the basic working diapason) and $R=1.0 ~m$ when
working inside the $70 ~m$ diapason.  In each separate event the
contribution of energy flux through each chamber to the energy
flux for each section of distances was taken into account. The
measured part of the EAS spectrum was also divided into small
intervals with respect to the $N_e$ (providing the optimal
statistics for each of the intervals). For each interval of $N_e$
the mean value (over all events of this interval) of the energy
flux  density at the given distance from the EAS axis was derived.
The result was multiplied by the area of $R$ cm width ring of the
mean radius equal to given distance. After this procedure it
becomes possible to estimate the flux of hadron energy for each
interval of $N_e$  within the desired intervals of core distances.

The whole experimental material was processed in the computing
centers of the Institute of Physics of the Geogian Academy of
Sciences, of the Ministry of Communications of Georgia and in the
computer firm Mantex ltd. (Tbilisi).

As compared with \cite{2,5} the body of processed statistical
material was increased essentially and further improvement of
algorithms and data handling methods was carried out. As well as
increased precision  this gave us the possibility to widen the
diapason of $N_e$  under investigation.

The data of the different runs of experiment were analyzed. These
runs differed by configuration of calorimeter (number of layers
and presence or absence of target), by triggering conditions and
by criteria of preliminary selection of events. As it was said
above, at different runs of the experiment calorimeter contained
from 16 to 19 layers of ionization chambers. Above the
calorimeter was placed or was absent the carbon target which was
800 $g/cm^2$  thick. The triggering conditions also might differ
in different runs. In accordance with this the whole analyzed
material may be naturally divided in parts, in frames of which
the configuration of the calorimeter, triggering criteria and
selection conditions remained stable. The triggering conditions
and conditions of event registration are described in detail in
\cite{7}.

The events registered at different configurations of installation
and at different triggering conditions were selected according to
the following conditions:

a) One part of the used archive bank contains the events which
were subjected to preliminary selection at time of the creation of
the bank. The requirements of selection were: the density of
electrons at the distance $73 ~m$ from the center of installation
greater then 0.45 particles/$m^2$~; X,Y coordinates of EAS axis
intersection with the plane of the central scintillator carpet
less then $7.0 ~m$; EAS age parameter $0.4 < s < 1.6$; total
number of particles in EAS  $N_e  > 1.3*10^5$ .  By reprocessing
of these data under the new conditions a part of the calculated
parameters shifted in one or another direction, often by a rather
large factor.

b) Another part of the bank is constituated of events which were
not subjected to the preliminary selection procedure. According to
the conditions of recent processing of the data the selected
events would satisfy the following requirements: at least one
particle must be registered in each group of scintillators at
distances 15, 20 and $73 ~m$ and in the central carpet; X,Y
coordinates of EAS axis intersection with the plane of the central
scintillator carpet less then $100 ~m$; age parameter $0.3 < s <
1.9$. Namely under these conditions were reprocessed the events of
"a)" above as well.  Criterion "b)" is much more liberal then
criterion "a)" however the data handling process itself is carried
out with the maximum accuracy achievable to-day.

At the final stage of analysis an additional restrictions were
put on the diapason of distances, on age parameter  $s$   and on
the precision and reliability of fitting of all basic parameters
in each separate event.

In such a way the analysis was carried out using the criteria
"a)"  and  "b)" for the 16-layer calorimeter without target  and
using criterion "b)" for the 19-layer calorimeter with the target
and without it. Moreover, the statistically greater material,
obtained by means of the 16-layer calorimeter without the target,
was divided into two parts of almost equal bodies with different
diapasons of registration zenith angles of 0 - 20 and 20 - 30
degrees. The aim was the same - to check the presence and
localization of peculiarities of $E_h(N_e)/N_e$   behavior at
different zenith angles of event registration.

200 000 events were selected out of the whole body of bank - 350
000 events, according the above quoted criteria. In the final
results  51 000 events contributed due to screening by the
conditions of the diapason of distances, of age parameter and of
EAS parameters fitting the accuracy requirement needed.

\section{\bf Experimental results  and discussion}

~~~The specific energies $E_h(N_e)/N_e$   of EAS hadronic
component for the above quoted configurations of installation and
selection conditions are given in Fig.1:

curve 1 -   16 layer calorimeter, without target. The material
not subjected to preliminary selection  was handled according
to criterion "b)",   37000 events in total. The diapason of
registration zenith angles 0 -30 degrees.

curve 2 -   16 layer calorimeter, without target. The
material subjected to preliminary selection,  was then handled
according criterion "b)",   8 000 events in total. The diapason
of registration zenith angles 0 - 30 degrees.

 curve 3 -   19 layer calorimeter, with target.
%
\begin{figure}[ht]
\centering
\begin{picture}(70,150)(140,240)
\includegraphics*[scale=.6]{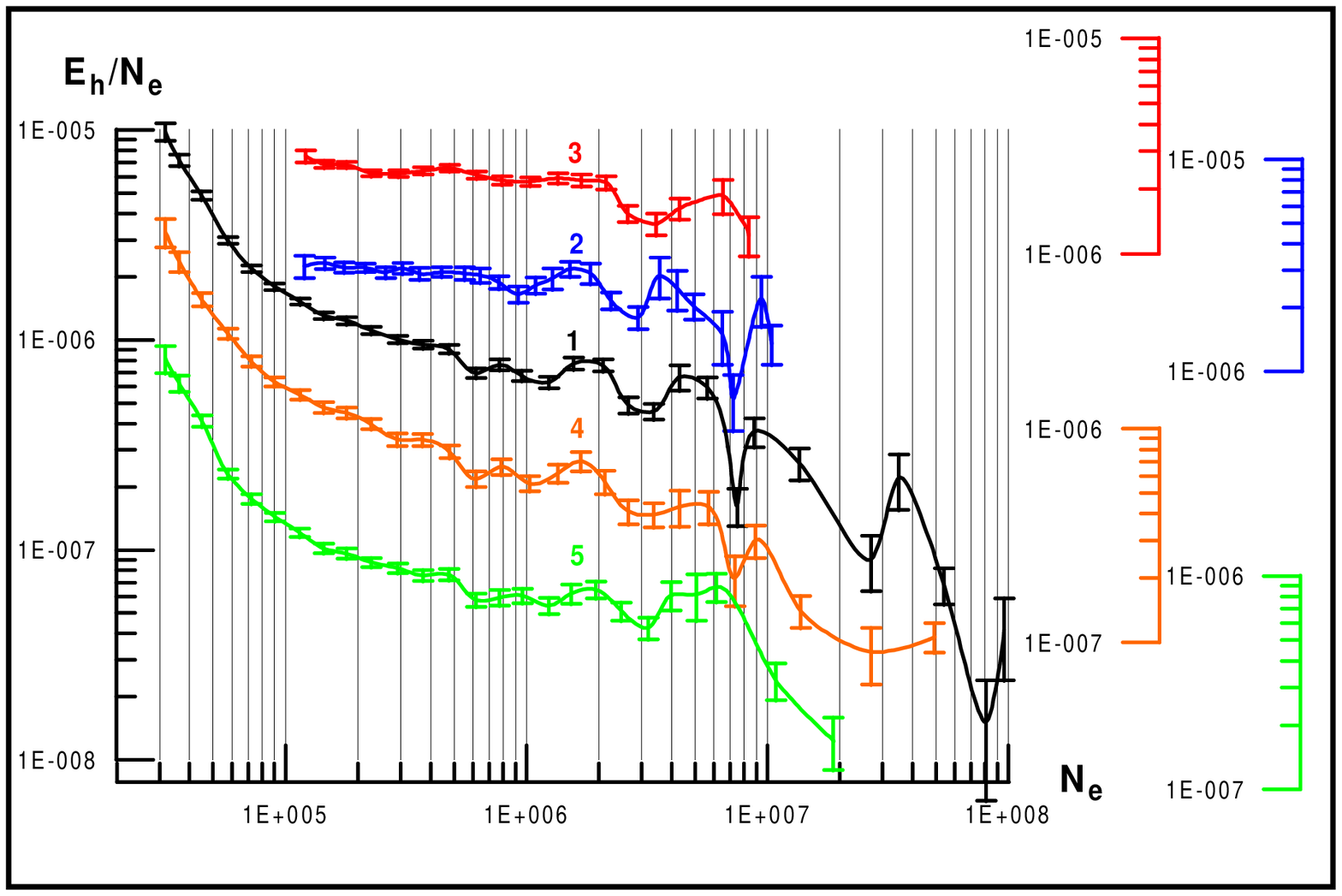}
\end{picture}

\parbox{10cm}{\vspace{10mm}{\bf Fig.1~:}~~ The specific energies
$E_h(N_e)/N_e$   of EAS hadronic component for the different
configurations of apparatus and selection conditions. Details are
given in the text.  The left vertical scale (arbitrary units) is
given for curve 1. The other curves are slightly shifted up or
down to separate them from one another and to clarify the
picture. The corresponding parts of vertical scales are shown at
the right edge of the figure. }
\end{figure}
The material  subjected to preliminary selection, was handled
according to criterion "b)", 6 000 events in total. The diapason
of registration zenith angles 0 - 30 degrees.

curves 4 and 5 - the material constituating the curve 1 is divided
in two parts according to the diapason of registration zenith
angles -   0 - 20 degrees (19 000 events) and 20 - 30 degrees (18
000 events) correspondingly.

The lower borders by $N_e$  are determined by the material
available  to-day and by triggering conditions (curves 1,~4 and 5)
or by conditions of preliminary selection (curves 2 and 3). The
upper borders are determined by statistical restrictions due to
high steepness  of EAS spectrum $I(N_e)$. The error bars allow for
statistical errors. The errors of $N_e$ determination are of the
order of 10 \%. The difference of mean slopes of the curves to the
left from the values $N_e < 10^6$ is easy to explain by the
difference in triggering conditions $E_h(N_e)/N_e$ and of
selection criteria. The solid curves - splain, drawn through the
experimental points.

 In the  region of $N_e  > 5*10^5$ the deeps on dependence
$E_h(N_e)/N_e$ are observed. Most clearly these irregularities are
revealed in the region $N_e  > 2*10^6$~, where the identical
localization of these phenomena on all shown curves may be easily
traced. The indicated deeps in $E_h(N_e)/N_e$ dependence are
located in the regions of the same values of $N_e$ , regardless of
experimental material used and of triggering and selection
conditions (curves 1,~2 and 3). Moreover, the localization of
these deeps does not depend on the zenith angle of  event
registration (curves 4 and 5).

In the region of relatively low values of $N_e$ , namely at
~$5*10^5 < N_e  < 2*10^6$  the mutually equivalent irregularities
may be seen as well. Here however the further increase of
accuracy and of statistics is required. In the region $N_e  <
5*10^5$  at this stage of investigation the $E_h(N_e)/N_e$
dependence behavior reveals as completely smooth.

Actually here we are dealing with the results, obtained in
several independent experiments. So at the  given stage of data
processing one can regard the existence of irregularities on
$E_h(N_e)/N_e$   dependence as reliably established one.

Apparently the nice possible explanation for the creation of such
behavior of EAS hadronic component flux can be given if one
supposes  the existence of PCR nuclei spectra cutoff in the
primary energy region $>10^{16}$  eV. The destruction of PCR
nuclei is caused by their interactions with some light particle
(component of nonbaryonic dark matter~$\!$?) forming the
monochromatic low temperature background in the interstellar
space \cite{1,3}.

According to this picture the cutoffs of PCR nuclei spectra have a
threshold character and depend on the primary energy of the
nucleus and on the energetic threshold of its destruction. For the
heavy nuclei of Fe group this phenomenon takes place at $N_e$ -s
higher then $1.5*10^7$ , which correspond to primary energies of
the order of $2.6* 10^{16}$  eV. The deeps at the $N_e$ values
higher than $2*10^6$ and $6*10^6$ are reflecting the destruction
of He group and of middle group nuclei. At the energies higher
then $5*10^{16}$, at which the nuclei of Pb group are destructed
(the last deep on the curve 1), the protons are dominating in the
composition of PCR, may be with exception of small part of
survived nuclei and of the most heavy nuclei like Uranium. The
latter are destructed at the energies of the order of $10^{17}$
eV.

\section{\bf The mechanism of the irregularities formation}

~~~The essence of the deep formation mechanism on $E_h(N_e)/N_e$
dependence is as follows.

At a fixed value $E_0$ of primary particle energy the number $N_e$ of
particles in EAS fluctuates within a rather wide range with the
certain mean value $N_e$' , and vice versa, the certain value of
$N_e$ can be registered in EAS initiated by the primary particle
of the energy wedged within a rather wide interval of primary
energies with the certain mean value E'. There is some balance of
contributions in showers of the given $N_e$ from primary particles
of different energies,  depending on the slope of their spectrum.

~Let the  spectrum of the given certain component of PCR have the
form $I(E) = K^{-r}$.  Now let the spectrum of a given component
of PCR be cut off above some fixed value of primary energy $E^*$
(cutoff energy), i.e. at primary energies higher then E* the flux
of  this component of PCR sharply decreases (the value of
coefficient $K$ falls sharply).  It is clear that the spectrum of
the corresponding EAS will not be cut off above the value $N_e'$
since the showers of the sizes $N_e  > N_e'$  will still be
registered due to fluctuations of $N_e$  in the showers of total
energy $E_0 < E^*$. The showers initiated by the primary particles
(of the PCR component under consideration) of the energy $E_0 >
E^*$ will be present in extent of the new value of coefficient
$K$. It is established that $E_h$ rises in average with the rise
of $E_0$ (and thus with the rise of $N_e$).  Besides, at the fixed
energy $E_0$ the anticorrelation of $E_h$ and $N_e$ fits naturally
with the account of the approximate conservation of the sum of
$e$- and $h$-component energies. It follows here from, that in the
case of spectrum cutoff the flux of hadron  energy in EAS of $N_e
> N_e$* will decrease in average as compared with the case of the
absence of cutoff, since the EAS for such $N_e$-s  now will be
generated by the primary particles of the energies $E_0 < E^*$.
The balance of contributions in showers of some given $N_e$
generated by primary particles of different energies will be
violated. The width of the interval in which this effect does
occur will be of the order of width of the fluctuations of $N_e$ .
At higher $N_e$-s exceeding the frames of this interval, all
manifestations of the flux of this component will decrease
sharply. However above the indicated interval the balance of
contributions restores again and the mean energy of  hadrons in
EAS reaches its previous level. In the case of complete cutoff of
the spectrum EAS from this component will vanish completely. For
the mixed composition of PCR the spectra of different components
will be cut off at different $E^*$ -s  \cite{3} and one has to
expect deeps on  the $E_h(N_e)/N_e$   plot near the corresponding
values of $N_e$ - s. The widths of these deeps are determined by
the ranges of $N_e$ fluctuations.

This mechanism of   $E_h(N_e)/N_e$   dependence formation is
analyzed in detail  in  \cite{3}, where the principles and
results of the calculation of its expected features are given. It
is to be noticed however, that the level of data processing and
analysis in \cite{3} was lower then reached  to day, so the
numerical values of model parameters and the localization  of the
signals along the $N_e$ axis differ remarkably from the present
results. New parameters, corresponding to recent level of data
handling are in preparation and will be published soon.

\vspace{5mm}

\noindent {\bf  ACKNOWLEDGMENTS} \vspace{3mm}

The authors express their deep gratitude to O.V.Kancheli for
numerous fruitful discussions and advise. The authors are
sincerely grateful  to Yu.D.Kotov for discussion, V.P.Pavluchenko
and V.I.Iakovlev for useful discussions concerning the data bank
and installation, N.R Tkeshelashvili, D.R .Tomaradze and
A.G.Mulkidjanian  -  the members of the staff of the Ministry of
communications of Georgia -  for their assistance in primary
treating of the data array, J.M.Henderson - for  interest to this
investigation and his assistance in preparation of this paper.

This investigation is in part supported by a grant from the
Georgian  Academy  of  Sciences and by the firm Mantex~ltd.
(Tbilisi).


\end{document}